\documentclass[prd,twocolumn,tightenlines,preprintnumbers,
showpacs,superscriptaddress,nofootinbib,amsmath,amssymb]{revtex4}
\usepackage{bm}

\begin{document}

\preprint{CECS-PHY-06/22}
\preprint{hep-th/0611202}

\title{Surveying the Three-Dimensional Fixed Points of T-Duality}

\author{Eloy Ay\'on--Beato}
\email{ayon-at-cecs.cl}
\affiliation{Centro~de~Estudios~Cient\'{\i}ficos~(CECS),%
~Casilla~1469,~Valdivia,~Chile.}
\affiliation{Departamento~de~F\'{\i}sica,~CINVESTAV--IPN,%
~Apdo.~Postal~14--740,~07000,~M\'exico~D.F.,~M\'exico.}
\author{Gast\'on Giribet}
\email{gaston-at-df.uba.ar}
\affiliation{Departamento de F\'{\i}sica, Universidad de Buenos Aires,
Ciudad Universitaria, Pabellon I (1428), Buenos Aires, Argentina.}

\begin{abstract}
We explore the family of fixed points of T-Duality transformations
in three dimensions. For the simplest nontrivial self-duality
conditions it is possible to show that, additionally to the
spacelike isometry in which the T-Duality transformation is
performed, these backgrounds must be necessarily stationary. This
allows to prove that, for nontrivial string coupling, the low energy
bosonic string backgrounds which are additionally self-T-dual along
an isometry direction generated by a constant norm Killing vector
are uniquely described by a two-parametric class, including only
three nonsingular cases: the charged black string, the exact
gravitational wave propagating along the extremal black string, and
the flat space with a linear dilaton. Besides, for constant string
coupling, the only self-T-dual lower energy string background under
the same assumptions corresponds to the Coussaert--Henneaux
spacetime. Thus, we identify minimum criteria that yield a
classification of these quoted examples and only these. All these
T-dual fixed points describe exact backgrounds of string theory.
\end{abstract}


\maketitle

\section{\label{sec:Int}Introduction}

One of the most fascinating aspects of string theory is the
existence of the duality symmetries, which turn out to relate
different regimes of the theory that, a priori, would seem to be
substantially distinct. The catalog of duality symmetries playing an
important role in string theory includes the quoted T-duality, which
relates different spacetime configurations, so that suggesting that
a more satisfactory picture is achieved when one considers those
spacetimes as describing diverse sectors of the same \emph{string
model}. For instance, duality transformations may map flat
backgrounds in curved ones, with the cost of generating a nontrivial
antisymmetric tensor. Moreover, even the topological aspects of the
space may abdicate in presence of T-duality transformations.
Consequently, such duality symmetry can be thought of as manifesting
a deeper relation existing between all the fields arising in the
formulation of the theory. In general, the understanding of such a
duality between two backgrounds is that both are, actually, two
spacetime interpretations of the same string solution. A particular
realization of this idea is the fact that, even though a given low
energy string solution and its dual can correspond to a pair of
distinct background configurations, the conformal field theory
description of both models is indeed the same \cite{Rocek:1991ps}
and, thus, both are equivalent from the string theory point of view
\cite{Dijkgraaf:1991ba}.

The interpretation of T-duality as a symmetry manifesting such a
deep aspect of string theory can be found with particular
conciseness in Refs.~\cite{Tseytlin:1994sb,Horowitz:1994ei}. This
perspective establishes that the spacetime metric provides just a
crude description of what string geometry actually is. Then, the
metric would not be a proper characterization of the \emph{string
geometry} since it is not enough to fully realize the symmetries of
the string theory. To be precise, fundamental properties of the
metric as its curvature, its causal structure, or its asymptotic
behavior are not, in general, invariant under duality
transformations \cite{Horowitz:1994ei}. Furthermore, it is currently
suggested that the existence of T-duality could be seen as a
manifestation of the fact that the spacetime itself is merely an
emergent notion rather than a fundamental element in string theory
\cite{Seiberg:2006wf}. Then, this point can be regarded as one of
the motivations to study the properties remaining invariant in the
space of solutions under duality and, in particular, to study the
fixed points of it. The physics at the fixed points of T-duality
manifests important properties of the string theory that, in other
regimes, could remain hidden. Such aspects include the well known
enhancement of the symmetries of the theory, with which we are
familiarized due to the well known case of string theory on flat
space compactified on the self-dual radius $R=\sqrt{2\alpha'}$.
Besides, the thermodynamics of the self-dual solutions presenting
event horizons also exhibits particular properties
\cite{Horowitz:1993wt}, and it can be also thought of as an
additional motivation to study it within the context of string
theory in curved spacetime. More importantly, self-T-dual
backgrounds are believed to play an important role in the search for
string theory vacua \cite{Faraggi:2004xf}.

Here we will analyze T-duality within the context of the
three-dimensional low energy effective theory coming from the
bosonic string theory. A particular property of three dimensions is
that all the low energy solutions with constant string coupling,
i.e.\ constant dilaton, turn out to be exact solutions to the string
theory beyond the field theory approximation. This is related to the
fact that, in three dimensions, the Weyl tensor is identically zero
and thus the Riemann tensor is completely determined by the Ricci
tensor through
$R_{\mu\nu\rho\sigma}=g_{\mu\rho}R_{\nu\sigma}+g_{\nu\sigma}R_{\mu\rho}
-g_{\nu\rho}R_{\mu\sigma}-g_{\mu\sigma}R_{\nu\rho}
-\frac{1}{2}(g_{\mu\rho}g_{\nu\sigma}-g_{\mu\sigma}g_{\nu\rho})R$.
Besides, in three dimensions the second order Gauss--Bonnet term
vanishes, $R_{\mu\nu\rho\sigma}R^{\mu\nu\rho\sigma}+R^{2}
-4R_{\mu\nu}R^{\mu\nu}=0$, and the antisymmetric Kalb--Ramond field
strength $H_{\mu\nu\rho}$ turns out to be proportional to the volume
three-form $\eta_{\mu\nu\rho}$. All these conditions enable one to
make use of the field redefinition freedom and eventually show that
a picture in which the $D=3$ effective action has all
$\alpha^{\prime}$ corrections depending only on the derivatives of
the dilaton does exist \cite{Horowitz:1994ei}. Then, those solutions
of the three-dimensional low energy effective theory for which the
dilaton results to be constant remain exact to all orders. Moreover,
all the solutions to Einstein theory with negative cosmological
constant, $\Lambda<0$, can be actually embedded in the
three-dimensional string theory. In fact, the leading order solution
with constant string coupling, i.e.\ constant dilaton, in three
dimensions turns out to be the three-dimensional Anti-de Sitter
space, or its cosets over discrete subgroups. This includes, of
course, the case of the Ba\~{n}ados--Teitelboim--Zanelli black hole
\cite{Banados:wn,Banados:1992gq} and the Coussaert--Henneaux
\cite{Coussaert:1994tu} geometry (see also \cite{Clement:1992ke}).
The last case is specially remarkable by its self-T-dual character
\cite{Coussaert:1994tu}. For the case of nonconstant dilaton,
solutions which do not receive corrections do exist as well; the
charged black string is perhaps the most celebrated example of this
\cite{Horne:1991cn}. Thus, the catalog of solutions of
three-dimensional string theory is rich enough to consider it as an
interesting toy model to study the physical content of the theory.

In the next section we review the T-duality transformations and
analyze the simplest properties that a nontrivial string background
must satisfy in order to be a fixed point of these transformations;
the possibility of having more general self-dual backgrounds is
discussed in Appendix C. In Sec.~\ref{sec:Birk} we establish a
Birkhoff's theorem for such spacetimes of the simplest nontrivial
class satisfying the low energy string equations. In doing this, we
find two stationary branches depending on whether a nontrivial
string coupling is allowed or is not. For a trivial string coupling,
we show that the only self-T-dual background with a constant dilaton
is precisely the previously quoted Coussaert--Henneaux spacetime
\cite{Coussaert:1994tu}. This is actually related to the fact that
the leading order three-dimensional solutions to the effective
equations with constant dilaton must be constant negative curvature
spacetimes as was mentioned before, i.e.\ they are locally $AdS_3$.
In the other branch, corresponding to backgrounds exhibiting a
nonconstant dilaton, it is proven that the simplest nontrivial
self-T-duality configurations are uniquely described by a
two-parametric family. This class is analyzed separately in
Sec.~\ref{sec:class} using as guide the existence of hypersurface
orthogonal Killing fields. The above allows a classification
depending on the values of the integration constants and containing
just three self-T-dual backgrounds which are regular or at least no
contains naked singularities. For generic values of the integration
constants the backgrounds correspond to the charged black string
\cite{Horne:1991gn}; for an extremal case they become an exact
gravitational wave propagating along the extremal black string
\cite{Garfinkle:1992zj}, and for vanishing constants we recover flat
space with a linear dilaton. Some relevant information is included
as Appendices. The first Appendix contains the explicit form of the
lower energy string equations for the self-T-dual configurations
under study. A second one is devoted to justify the gauge elections
made on the dilaton in Sec.~\ref{sec:Birk}. As it was mentioned, a
third Appendix explores the possibility of constructing more general
self-dual backgrounds besides the ones analyzed in the main part of
the paper.

\section{\label{sec:sTd}On self-T-duality}

The low energy effective action, describing the particle limit of the
bosonic string dynamics in $2+1$ dimensions, is given by
\begin{eqnarray}
S &=& \int{d^{3}x}\sqrt{-g}e^{-2\Phi}\left( \frac{4}{k}+R
+4\nabla_{\alpha}\Phi\nabla^{\alpha}\Phi\right. \nonumber \\
  & & \qquad{}-\left.
\frac{1}{12}H_{\alpha\beta\gamma}H^{\alpha\beta\gamma} \right),
\label{eq:ac}
\end{eqnarray}
where $\Phi$ is the dilaton field, $H_{\alpha\beta\gamma}
=\partial_{[\alpha}B_{\beta\gamma]}$ is the field strength of the
Kalb--Ramond field $B_{\alpha\beta}$, and the constant $k$ turns out
to be related with the central charge of the sigma model. For those
cases where a given background has been recognized as an exact
conformal one, the value of $k$ typically corresponds to the level
of the WZNW model involved. The equations of motion yielding from
the action above read
\begin{subequations}
\label{eq:les}
\begin{eqnarray}
R_{\alpha\beta}+2\nabla_{\alpha}\nabla_{\beta}\Phi
-\frac{1}{4}H_{\alpha\mu\nu}{H_{\beta}}^{\mu\nu} &=&0, \label{eq:Ein}\\
\nabla_{\mu}(e^{-2\Phi}H^{\mu\alpha\beta})       &=&0, \label{eq:KR}\\
\Box\Phi-2\nabla_{\alpha}\Phi\nabla^{\alpha}\Phi
+\frac{2}{k}+\frac{1}{12}H_{\alpha\beta\gamma}H^{\alpha\beta\gamma}
&=&0.\label{eq:dil}
\end{eqnarray}
\end{subequations}

A very interesting property of string theory is the fact that the
above low energy string equations turn out to be invariant under the
often called T-duality transformations. This is a symmetry of the
theory that maps any given solution $(g_{\mu\nu},B_{\mu\nu},\Phi)$
with a translational symmetry, lets say along the direction
$\bm{\partial_x}$, to another solution
($\widetilde{g}_{\mu\nu},\widetilde{B}_{\mu\nu},\widetilde{\Phi}$)
given by the Buscher rules \cite{Buscher:1987qj}
\begin{eqnarray}
\widetilde{g}_{xx} &=&1/g_{xx}, \nonumber \\
\widetilde{g}_{xm} &=&B_{xm}/g_{xx}, \nonumber \\
\widetilde{g}_{mn} &=&g_{mn}-(g_{xm}g_{xn}-B_{xm}B_{xn})/g_{xx},
                      \nonumber\\
\widetilde{B}_{xm} &=&g_{xm}/g_{xx}, \nonumber \\
\widetilde{B}_{mn} &=&B_{mn}-2g_{x[m}B_{n]x}/g_{xx}, \nonumber \\
\widetilde{\Phi}   &=& \Phi -\frac{1}{2}\ln g_{xx},\label{eq:T-dual}
\end{eqnarray}
where $m,n$ represent the other directions, different from $x$.

Within this context, a natural question arises as to whether the
previous system allows the existence of self-T-dual configurations,
i.e.\ the existence of string backgrounds which result to be fixed
points of the above T-duality transformations. In fact, it is well
known that this is actually the case, as the study of the
Coussaert--Henneaux spacetime shows \cite{Coussaert:1994tu}, and
what we want to analyze here is the existence of other relevant
examples. These self-dual configurations are to be defined as those
having a special symmetric direction such that they coincide with
the corresponding T-dual modulo diffeomorphisms and gauge
transformations.

The simplest way to obtain a self-dual configuration is to fix
$g_{xx}=1$ and additionally to impose $g_{xm}=0$ and $\bm{B}=0$.
Under these conditions the background is manifestly self-T-dual.
However, we are interested in less trivial examples. That is why
here we will focus our attention on all the particular cases of
self-T-dual configurations that present a Killing vector
$\bm{m}=\bm{\partial_\varphi}$ with constant norm
$\bm{g}(\bm{m},\bm{m})=a^{2}=\mathrm{const}$; in this way the
special direction is given by an appropriated normalization
$\bm{\partial_x}=a^{-1}\bm{\partial_\varphi}$. Having $g_{xx}=1$
leaves invariant the dilaton and the component $g_{xx}$ itself in
the transformations (\ref{eq:T-dual}). A remaining condition in
order to obtain self-T-duality would be then $B_{xm}=g_{xm}$. Notice
that this last condition has to be understood as holding up to gauge
transformations and coordinate transformations preserving the
Killing direction $\bm{\partial_x}$. Further, we will use this
condition as a gauge fixing actually. Summarizing, demanding the
existence of a constant norm Killing field is the simplest way to
obtain nontrivial self-dual backgrounds. This is not a necessary
requirement, that is why we study in Appendix \ref{app:par} a
possibility of constructing backgrounds without such property.
Nevertheless, straightforward attempts to find an explicit example
do not succeed.

The above considerations motivate the study of those string
backgrounds for which the metric exhibits the generic form
\begin{equation}
\bm{g}=\bm{g}^{(2)}+a^{2}[\bm{d\varphi}+W_{m}(x^{n})\bm{dx}^{m}]^{2},
\label{eq:gTd}
\end{equation}
where $\bm{g}^{(2)}=g_{mn}^{(2)}(x^{n})\bm{dx}^{m}\bm{dx}^{n}$
($m,n=0,1$) is the metric of a two-dimensional spacetime. In the
case of stationary configurations we have an additional timelike
Killing field $\bm{k}=\bm{\partial_t}$, and the two-dimensional
metric can be written as
\begin{equation}
\bm{g}^{(2)}=-N(r)^{2}F(r)\bm{dt}^{2}+\frac{\bm{dr}^{2}}{F(r)},
\label{eq:g2}
\end{equation}
where a gauge election still remains to be fixed.

A particular example of the previous class is the metric of the
Coussaert--Henneaux spacetime, which is given by
\cite{Coussaert:1994tu,Ayon-Beato:2004if}
\begin{subequations}
\label{eq:CH}
\begin{eqnarray}
F(r) &=&1, \\
N(r) &=&\cosh(2r/l), \\
W_{m}(r) &=&\frac{1}{a}\sinh(2r/l)\delta_{m}^{t},
\end{eqnarray}
\end{subequations}
and where $\varphi$ is periodically identified with $\varphi+2\pi$.
This example describes a spacetime with constant negative curvature
$R=-6/l^{2}$ and isometry group $SO(2)\times{SO(2,1)}$. This
background is also a solution to low energy string theory equations
(\ref{eq:les}) with $\Phi=\mathrm{const.}$ and
$H_{\alpha\beta\gamma}$ given by the volume three-form modulo a
constant \cite{Coussaert:1994tu}. This is because, for a constant
dilaton, the low energy string action (\ref{eq:ac}) becomes the
Einstein-Hilbert action with an effective cosmological constant
$\Lambda=-1/l^{2}=-1/k$. This is which allows to embed any
three-dimensional constant negative curvature spacetime within low
energy string theory, as it has been explicitly shown in
Refs.~\cite{Horowitz:1993} and \cite{Coussaert:1994tu} for the BTZ
black hole \cite{Banados:wn,Banados:1992gq} and the
Coussaert--Henneaux spacetime \cite{Coussaert:1994tu}, respectively;
see also \cite{Horowitz:1994ei,Horowitz:1994rf}. These constant
curvature solutions could seem to be no so interesting from the
string theory point of view since them solve the equations of
motions in a regime where string theory just mimics General
Relativity. Nevertheless, it is worth noticing that such a regime
can be, in some cases, related to a nontrivial one by T-duality,
e.g. the BTZ black hole turns out to be T-dual to the charged black
string solution which presents a nontrivial dilaton configuration
\cite{Horowitz:1993}.

\section{\label{sec:Birk}Birkhoff's Theorem for a Class of
Self-T-Dual Backgrounds}

In this section we present a Birkhoff's theorem for the class of
three-dimensional string backgrounds which remain invariant under
T-duality transformations generated by constant norm directions. In
fact, we prove that additionally to the constant norm spacelike
Killing field $\bm{m}$ realizing T-duality, there exists another
Killing field $\bm{k}$ which turns out to be timelike; so that the
resulting three-dimensional self-T-dual configurations must be
necessarily stationary. This will allow us to identify an important
set of quoted self-T-dual examples within the class considered here.

In order to prove this result we will make use of the following
explicit form for the metric
\begin{eqnarray}
\bm{g} &=&-N(t,r)^{2}F(t,r)\bm{dt}^{2}+\frac{\bm{dr}^{2}}{F(t,r)}
\nonumber \\
       & &\qquad\qquad\qquad{}
          +a^{2}\left(\bm{d\varphi}+W(t,r)\bm{dt}\right)^{2},
\label{eq:gexpl}
\end{eqnarray}
where we have already used part of the diffeomorphism invariance in
order to fix the components $g_{tr}=0=g_{\varphi{r}}$. The low
energy string equations (\ref{eq:les}), once evaluated for the
ansatz (\ref{eq:gexpl}) assuming that the dilaton and the
Kalb--Ramond field are also $\varphi$-independent, are explicitly
written down in Appendix \ref{app:SEqs}. Their integration depends
on a remaining gauge election. Such gauge election and the resulting
local solutions depend on the nature of the surfaces
$e^{-\Phi}=\mathrm{const.}$, defined by the string coupling; these
surfaces can be timelike, spacelike or null, or they can be not
defined at all if the string coupling turns out to be a constant.
Hence, the corresponding open neighborhoods can be classified
according to the three possible signs of the following norm
\begin{equation}
\nu^{2}\equiv\nabla_{\mu}e^{-\Phi}\nabla^{\mu}e^{-\Phi}
=Fe^{-2\Phi}\left((\Phi^{\prime})^{2}
-\frac{(\dot{\Phi})^{2}}{N^{2}F^{2}}\right). \label{eq:norm}
\end{equation}
For each sign we make a separate analysis employing a different
choice of coordinates, which also make easier the integration of the
field equations. Notice that it is sufficient to analyze the case of
positive $F$ only, since changing the sign of $F$ just corresponds
to changing the sign of $\nu^{2}$.

\subsection{\label{subsec:spacelike}Case $\protect\nu^2>0$: regions
$r<r_{-}$ or $r>r_{+}$}

For open neighborhoods where
$\nabla_{\mu}{e^{-\Phi}}\nabla^{\mu}{e^{-\Phi}}>0$ the surfaces
$e^{-\Phi}=\mathrm{const.}$ are timelike and we can fix the gauge
such that
\begin{equation}
e^{-\Phi (t,r)}=\frac{r}{a}, \label{eq:gaugePhi(r)}
\end{equation}
i.e.\ in this coordinates the string coupling $e^{\Phi}$ vanishes
when $r$ goes to infinity. We provide a complete justification for
this election on Appendix \ref{app:gauge}. With this choice, the low
energy string equations (see Appendix \ref{app:SEqs}) can be easily
integrated, yielding the following metric functions
\begin{subequations}
\begin{eqnarray}
F(t,r) &=& \frac{r^{2}}{k}-M+\frac{J^{2}}{4r^{2}}, \label{eq:Fgr}\\
N(t,r) &=& \frac{a}{r}N_{0}(t), \label{eq:Ngr} \\
W(t,r) &=&-\frac{J}{2r^{2}}N_{0}(t)+W_{0}(t), \label{eq:Wgr}
\end{eqnarray}
\end{subequations}
where $N_{0}$ and $W_{0}$ are arbitrary functions of $t$, and where
$M$ and $J$ are two integration constants. Additionally, we obtain
the self-duality condition
\begin{equation}
c^{2}=\frac{J^{2}}{a^{4}}, \label{eq:c2J}
\end{equation}
where the constant $c$ is related to the axion charge per unit
length, see Eq.~(\ref{eq:solH}) in the Appendix \ref{app:SEqs}.

Moreover, functions $N_{0}$ and $W_{0}$ above can be eliminated
by the following coordinate transformation
\begin{equation}
\textstyle(t,r,\varphi) \mapsto\left(\int{N_{0}(t)\mathrm{d}t},
\;r,\;\varphi+\int{W_{0}(t)\mathrm{d}t}\right) , \label{eq:sctshp}
\end{equation}
which makes evident that there is no time dependence.

Next, we have to integrate the Kalb--Ramond field from
$\bm{H}=\bm{dB}$ and Eq.~(\ref{eq:solH}), considering that the
antisymmetric field can be determined only up to gauge
transformations $\bm{B}\rightarrow\bm{B}+\bm{dA}$, where $\bm{A}$ is
an arbitrary vector field. Then, we end up with the following
stationary configuration
\begin{subequations}
\label{eq:sol}
\begin{eqnarray}
\bm{g} &=&-\frac{a^{2}}{r^{2}}
\left(\frac{r^{2}}{k}-M+\frac{J^{2}}{4r^{2}}\right) \bm{dt}^{2}
\nonumber \\
       & &{}+\left(\frac{r^{2}}{k}-M+\frac{J^{2}}{4r^{2}}\right)^{-1}
       \bm{dr}^{2}
\nonumber \\
       & &{}+a^{2}\left(\bm{d\varphi}-\frac{J}{2r^{2}}\bm{dt}\right)^{2},
\label{eq:solg} \\
\Phi   &=&-\ln \left( \frac{r}{a}\right) , \label{eq:solPhi} \\
\bm{B} &=&\frac{Ja^{2}}{2r^{2}}\bm{dt}\wedge\bm{d\varphi}.
\label{eq:solB}
\end{eqnarray}
\end{subequations}
According to our gauge election this solution is valid only in the
regions where $F>0$. For values of the integration constants such
that $|J|\leq {M\sqrt{k}}$ the function $F$ turns out to be positive
for $r<r_{-}$ or $r>r_{+}$, where $r_{\pm}$ are the positive roots
of the equation $F=0$, and these are defined by
\begin{equation}
M=\frac{r_{+}^{2}+r_{-}^{2}}{k},\qquad
|J|=\frac{2\,r_{+}r_{-}}{\sqrt{k}}. \label{eq:MJ2rpm}
\end{equation}
On the other hand, for values of the integration constants obeying
$|J|>M\sqrt{k}$ the function $F$ turns out to be positive
everywhere. Accordingly, the coordinate $r$ is restricted to be defined
within the region $\{r<r_{-}\}\cup \{r>r_{+}\}$ for
$|J|\leq{M\sqrt{k}}$ and has no restriction at all when
$|J|>M\sqrt{k}$.

It is easy to check that in the special direction
$\bm{\partial_x}=a^{-1}\bm{\partial_\varphi}$ not only $g_{xx}=1$
but additionally
\begin{equation}
B_{xm}=g_{xm}=-\frac{Ja}{2r^{2}}\delta_{m}^{t}, \label{eq:sd}
\end{equation}
and, as was mentioned at the beginning, this guarantees the
self-T-duality of the previous string background.

\subsection{\label{subsec:timelike}Case $\protect\nu^2<0$: region
$r_{-}<r<r_{+}$}

Now, let us move to the open neighborhoods where $\nu^{2}$ takes
negative values. In these regions the surfaces
$e^{-\Phi}=\mathrm{const.}$ are spacelike and the time
coordinate can be identified as (see Appendix \ref{app:gauge}),
\begin{equation}
e^{-\Phi(t,r)}=\frac{t}{a}. \label{eq:gaugePhi(t)}
\end{equation}
In this gauge the string coupling does decay in time. Then, as it
can be deduced from the equations in Appendix \ref{app:SEqs}, this
fact implies the following form for the metric functions,
\begin{subequations}
\begin{eqnarray}
F(t,r) &=&\frac{t^{2}}{a^{2}N_{0}(r)^{2}f(t)},  \label{eq:Fgt}\\
N(t,r) &=&-\frac{a}{t}N_{0}(r),  \label{eq:Ngt} \\
W(t,r) &=&-\frac{J\int{N_{0}(r)\mathrm{d}r}}{t^{3}}+W_{0}(t),
\label{eq:Wgt}
\end{eqnarray}
\end{subequations}
where $N_{0}$ and $W_{0}$ are arbitrary functions of $r$ and $t$,
respectively, and where
\begin{equation}
f(t)=-\frac{t^{2}}{k}+M-\frac{J^{2}}{4t^{2}},  \label{eq:f(t)}
\end{equation}
with $M$ and $J$ being two integration constants. In this case the
self-duality condition (\ref{eq:c2J}) is again fulfilled.

In order to preserve the condition $F>0$, the function $f$ must be
positive as well. Hence, the above solution turns out to be valid
only for $|J|\leq{M\sqrt{k}}$ within the range $t_{-}<t<t_{+}$,
where $t_{\pm}$ are the positive roots of the equation $f(t)=0$.

Once again, the functions $N_{0}$ and $W_{0}$ can be eliminated by
rescaling the coordinate $r$ and shifting the coordinate $\varphi$
appropriately. However, making the following coordinate
transformation
\begin{eqnarray}
&&(t,r,\varphi )\mapsto \nonumber \\
&&\textstyle\left( \int{N_{0}(r)\mathrm{d}r},t,\varphi+\int{W_{0}(t)
\mathrm{d}t}+\frac{J}{2t^{2}}\int{N_{0}(r)\mathrm{d}r}\right),\qquad~
\label{eq:t<->r}
\end{eqnarray}
which additionally interchanges the resulting coordinates $t$ and
$r$, the solution takes the same form as the background
(\ref{eq:sol}), but now restricted to satisfy the condition
${r^{2}}/{k}-M+{J^{2}}/{4r^{2}}<0$. This is obeyed only when
$|J|\leq{M\sqrt{k}}$ for $r_{-}<r<r_{+}$. The open regions analyzed
here for $|J|\leq{M\sqrt{k}}$ are naturally glued with the
corresponding ones of the previous section at the points
$r_{\pm}$ where $\nu^2$ changes its sign due to the vanishing of the
function $F$, see definition (\ref{eq:norm}).

Taking into account the discussion of this subsection and that of
the previous one, we can conclude that the union of the different
cases studied up to now gives rise to all the different patches of
the background (\ref{eq:sol}) for any value of its integration
constants $M$ and $J$. It remains to study now the case for which
$\nu^2$ vanishes in an open region.

\subsection{\label{subsec:null} Case $\protect\nu^{2}=0$:
Coussaert--Henneaux spacetime}

Then, let us close this section discussing the case for which $\nu$
vanishes in an open neighborhood. The fact that the surfaces
$e^{-\Phi}=\mathrm{const.}$ are null surfaces implies
\begin{equation}
\dot{\Phi}=FN\Phi^{\prime}.  \label{eq:subsPt}
\end{equation}
Inserting this condition in the dilaton equation (\ref{eq:dila}) we
obtain
\begin{equation}
c^{2}e^{4\Phi}=\frac{4}{k},  \label{eq:cCH}
\end{equation}
i.e.\ the dilaton must take a constant value $\Phi=\Phi_{0}$. This
implies that we must take a different gauge election in this case.
Then, we choose one for which the coordinate $r$ is a proper
distance: $F(t,r)=1$. The remaining metric functions are easily
determined from the nontrivial field equations (see Appendix
\ref{app:SEqs}) leading to the expressions
\begin{subequations}
\begin{eqnarray}
N(t,r) &=&N_{0}(t)\cosh\left(\frac{2r}{\sqrt{k}}+H_{0}(t)\right), \\
W(t,r) &=&\frac{N_{0}(t)}{a}\sinh
\left(\frac{2r}{\sqrt{k}}+H_{0}(t)\right) +W_{0}(t),\quad~
\end{eqnarray}
\end{subequations}
where $N_{0}$, $W_{0}$, and $H_{0}$ are time dependent integration
functions.

The functions $N_{0}$ and $W_{0}$ can be again eliminated by an
appropriate rescaling of the time $t$ and a shifting of the
coordinate $\varphi$, similarly as the ones used in the
transformation (\ref{eq:sctshp}). This is equivalent to choose
$N_{0}=1$ and $W_{0}=0$. The conditions for making the function
$H_{0}$ to vanish turns out to be more subtle. In
Ref.~\cite{Ayon-Beato:2004if} a Birkhoff's theorem was proven for
the class of metrics (\ref{eq:gexpl}) described by gravity in
presence of a negative cosmological constant. In particular, it was
proven that for a metric determined by the functions above there
exists a coordinate system in which $H_{0}=0$. The involved
transformation is highly nontrivial, however, it can be found for
any nontrivial function $H_{0}$, see the last Appendix of
Ref.~\cite{Ayon-Beato:2004if}. In terms of these coordinates the
final configuration reads
\begin{subequations}
\label{eq:solCH}
\begin{eqnarray}
\bm{g} &=& -\cosh^{2}\left(\frac{2r}{\sqrt{k}}\right)\bm{dt}^{2}
+\bm{dr}^{2}  \nonumber \\
&&{}+a^{2}\left[\bm{d\varphi}
+\frac{1}{a}\sinh\left(\frac{2r}{\sqrt{k}}\right)\bm{dt}\right]^{2},
\label{eq:gCH} \\
\Phi   &=&\Phi_{0}, \\
\bm{B} &=& -a\sinh\left(\frac{2r}{\sqrt{k}}\right)
\bm{dt}\wedge\bm{d\varphi},
\end{eqnarray}
\end{subequations}
where the metric (\ref{eq:gCH}) corresponds to the one of the
Coussaert--Henneaux spacetime with constant negative curvature
$R=-6/k$ \cite{Coussaert:1994tu}, if one imposes the identification
$\varphi=\varphi+2\pi$. The connection between the results of this
subsection and those of Ref.~\cite{Ayon-Beato:2004if} is not a mere
coincidence, it is due to the fact that, for
a constant dilaton, the three-dimensional low energy string effective
action becomes the Einstein--Hilbert term plus a negative
cosmological constant. Again, in the special direction
$\bm{\partial_x}=a^{-1}\bm{\partial_\varphi}$ the above background
satisfies $g_{xx}=1$ and
\begin{equation}
B_{xm}=g_{xm}=\sinh\left(\frac{2r}{\sqrt{k}}\right)\delta_{m}^{t},
\label{eq:sdCH}
\end{equation}
which manifestly shows its self-T-dual character.

\section{\label{sec:class}Classifying Backgrounds with Nontrivial
string coupling}

In the previous section it was shown that there are two
distinguishable branches of self-T-dual backgrounds with constant
norm self-dual direction, depending on whether the string coupling
is constant or is not. For a constant dilaton, the
Coussaert--Henneaux spacetime (\ref{eq:solCH}) was identified as
being the only possibility. On the other hand, for a nonconstant
dilaton, we obtained the general two-parameter solutions
(\ref{eq:sol}). In this section we will study these last solutions
in order to completely identify the different string backgrounds
contained within this class.

First, we notice that for $J=0$, $M>0$, the configuration
(\ref{eq:sol}) reduces to the uncharged black string
\cite{Horne:1991gn},
\begin{subequations}
\label{eq:ucbs}
\begin{eqnarray}
\bm{g}
&=&-\left(1-\frac{\mathcal{M}}{\hat{r}}\right)\bm{d\bar{t}}^{2}
+\left(1-\frac{\mathcal{M}}{\hat{r}}\right)^{-1}
\frac{k\,\bm{d\hat{r}}^{2}}{4\hat{r}^{2}}+\bm{dx}^{2},\qquad~
\label{eq:gucbs} \\
\Phi &=&-\frac{1}{2}\ln\left(\frac{\sqrt{k}\,\hat{r}}{a^{2}}\right),
\label{eq:Phiucbs}
\end{eqnarray}
\end{subequations}
with $\bm{B}=0$ and mass (per unit length)
$\mathcal{M}=r_{+}^{2}/\sqrt{k}$, where we used the coordinate change
\begin{equation}
(t,r,\varphi)\mapsto
\left(\bar{t}=at/\sqrt{k},\,\hat{r}=r^{2}/\sqrt{k},
\,x=a\varphi\right).
\end{equation}
It is known that this background corresponds to the direct product
of the two dimensional Witten black hole
\cite{Witten:1991yr,Perry:1993ry} and the line, which is evident
making the redefinition $\hat{r}=\mathcal{M}\cosh^2(\rho/\sqrt{k})$,
\begin{subequations}
\label{eq:Witten}
\begin{eqnarray}
\bm{g}
&=&-\tanh^{2}\left(\frac{\rho}{\sqrt{k}}\right)\bm{d\bar{t}}^{2}
+\bm{d\rho}^{2}+\bm{dx}^{2}, \label{eq:gWitten} \\
\Phi &=&-\ln\left[\frac{\sqrt{k^{1/2}\mathcal{M}}}{a}
\cosh\left(\frac{\rho}{\sqrt{k}}\right)\right]. \label{eq:PhiWitten}
\end{eqnarray}
\end{subequations}
This background is also known to be T-dual to the static BTZ
black hole \cite{Horowitz:1993}, which is an orbifold of the
$SL(2,\mathbb{R})_k$ WZNW model.

We will argue here that a similar situation occurs for $J\neq0$,
i.e.\ we will show that the configuration (\ref{eq:sol}) coincides
with the charged black string \cite{Horne:1991gn}, except for
special values of the integration constants. However, one may be
puzzled about the fact that both the uncharged black string solution
described above and the charged one turns out to be static
spacetimes, while the background (\ref{eq:sol}) does not seem to be
so. In order to clarify this point, we will make a classification of
the different spacetimes contained within the class (\ref{eq:sol})
in terms of their properties concerning the existence of
hypersurface-orthogonal Killing fields $\bm{k}^{\mathrm{s}}$. This
is equivalent to demand the fulfillment of the Frobenius
integrability condition
\begin{equation}
\bm{k}^{\mathrm{s}}\wedge \bm{dk}^{\mathrm{s}}=0,  \label{eq:Frob}
\end{equation}
for a given combination $\bm{k}^{\mathrm{s}}$ of the Killing fields.

On the one hand, for values such that $|J|\leq {M\sqrt{k}}$ we
observe that when $r_{+}>r_{-}$ the only combinations of the Killing
fields which are hypersurface-orthogonal and timelike in the
exterior region $r>r_{+}$ must be proportional to
\begin{equation}
\bm{k}^{\mathrm{s}}=\bm{\partial_t}+\frac{r_{-}}{r_{+}\sqrt{k}}
\bm{\partial_\varphi}.  \label{eq:ks}
\end{equation}
For the extremal case $r_{+}=r_{-}\neq0$ the hypersurface-orthogonal
Killing fields are also given by the previous expression, but they
become null in this limit. Finally, for $r_{+}=0=r_{-}$, any Killing
combination is hypersurface-orthogonal.

On the other hand, for $|J|\geq M\sqrt{k}$ there are no
hypersurface-orthogonal Killing fields if $J\neq0$, and the case of
vanishing $J$ allows both Killing fields to be
hypersurface-orthogonal. We will analyze each of these cases
separately in the following subsections, and we will show that they
actually describe spacetimes with different properties.

\subsection{\label{subsec:cbs}Case $r_+>r_-$: The charged black string}

For $r_{+}>r_{-}$ the existence of the stationary and
hypersurface-orthogonal Killing field (\ref{eq:ks}) in the exterior
region $r>r_{+}$ guarantees that this region is actually static,
i.e.\ the off-diagonal terms in metric (\ref{eq:solg}) are just an
artifact of the gauge that has been chosen. Then, we find convenient
to change to a new coordinate system adapted to
$\bm{k}^{\mathrm{s}}$ and where the staticity turns out to be
explicit; namely
\begin{eqnarray}
\left(t,r,\varphi\right) \; \mapsto \Biggl( \;
\hat{t} &=&\frac{a}{\sqrt{k}} \frac{(r_{+}t-r_{-}\sqrt{k}\,\varphi)}
{(r_{+}^{2}-r_{-}^{2})^{1/2}},  \nonumber \\
\hat{r} &=&\frac{r^{2}}{\sqrt{k}},  \nonumber \\
\hat{x} &=&\frac{a}{\sqrt{k}}\frac{(r_{+}\sqrt{k}\,\varphi-r_{-}t)}
{(r_{+}^{2}-r_{-}^{2})^{1/2}}\;\Biggr).
\end{eqnarray}
In these coordinates the string background (\ref{eq:sol}) takes the
form
\begin{subequations}
\label{eq:cbs}
\begin{eqnarray}
\bm{g} &=&-\left(1-\frac{\mathcal{M}}{\hat{r}}\right)\bm{d\hat{t}}^{2}
+\left(1-\frac{\mathcal{Q}^{2}}{\mathcal{M}\hat{r}}\right)
\bm{d\hat{x}}^{2}  \nonumber \\
&&{}+\left(1-\frac{\mathcal{M}}{\hat{r}}\right)^{-1}
\left(1-\frac{\mathcal{Q}^{2}}{\mathcal{M}\hat{r}}\right)^{-1}
\frac{k\,\bm{d\hat{r}}^{2}}{4\hat{r}^{2}}, \label{eq:gcbs} \\
\Phi &=&-\frac{1}{2}\ln\left(\frac{\sqrt{k}\,\hat{r}}{a^{2}}\right),
\label{eq:Phicbs} \\
\bm{B} &=&\frac{\mathcal{Q}}{\hat{r}}\bm{d\hat{t}}\wedge\bm{d\hat{x}},
\label{eq:Bcbs}
\end{eqnarray}
\end{subequations}
where $\mathcal{M}=r_{+}^{2}/\sqrt{k}$ and
$|\mathcal{Q}|=r_{+}r_{-}/\sqrt{k}$. This corresponds to the
three-dimensional charged black string \cite{Horne:1991gn} with mass
(per unit length) $\mathcal{M}$ and axion charge (per unit length)
$\mathcal{Q}$. This non-linear sigma model corresponds to the WZNW
model formulated on $SL(2,\mathbb{R})\times\mathbb{R}/\mathbb{R}$.

The T-dual properties of the charged black string were explored in
Refs.~\cite{Horne:1991cn,Horowitz:1993}. Its dualization along
$\hat{x}$ gives a boosted uncharged black string
\cite{Horne:1991cn}. Dualizing along a more general spacelike
direction both the BTZ black hole and another charged black strings
can be obtained \cite{Horowitz:1993}. As it has been seen here, the
charged black string is also a fixed point of the T-duality
transformation where the self-T-dual direction is given by
\begin{equation}
\bm{\partial_x}=\frac{1}{(\mathcal{M}^{2}-\mathcal{Q}^{2})^{1/2}}
\left(
\mathcal{M}\bm{\partial_{\hat{x}}}-|\mathcal{Q}|\bm{\partial_{\hat{t}}}
\right).  \label{eq:xcbs}
\end{equation}
Next, let us move to consider the extremal case.

\subsection{\label{subsec:gwebs}Case $r_+=r_-\neq0$: Gravitational wave
propagating along the extremal black string}

For the extremal case $r_+=r_-\neq0$ the hypersurface-orthogonal
Killing field (\ref{eq:ks}) turns out to be null. Such kind of null
symmetries are usually associated to the existence of gravitational
waves, and we will explicitly show that this is actually the case.
Using the following coordinates adapted to the null vector,
\begin{eqnarray}
\left(t,r,\varphi\right) \; \mapsto \biggl( \; v
&=&\frac{a}{2\sqrt{k}}(t+\sqrt{k}\,\varphi),  \nonumber \\
\hat{r} &=&\frac{r^{2}}{\sqrt{k}},  \nonumber \\
u &=&\frac{a}{\sqrt{k}}(t-\sqrt{k}\,\varphi)\quad\biggr),
\end{eqnarray}
it is possible to express the configuration (\ref{eq:sol}) as
\begin{subequations}
\label{eq:gwebs}
\begin{eqnarray}
\bm{g} &=& -
\left(1-\frac{\mathcal{M}}{\hat{r}}\right)2\,\bm{du}\bm{dv} +
\left(1-\frac{\mathcal{M}}{\hat{r}}\right)^{-2}
\frac{k\,\bm{d\hat{r}}^2}{4\hat{r}^2}\quad~  \nonumber \\
& & {} + \frac{\mathcal{M}}{\hat{r}}\bm{du}^2, \label{eq:ggwebs} \\
\Phi &=& -\frac12\ln\left(\frac{\sqrt{k}\,\hat{r}}{a^2}\right),
\label{eq:Phigwebs}\\
\bm{B} &=& \frac{\mathcal{M}}{\hat{r}}\bm{dv}\wedge\bm{du},
\label{eq:Bgwebs}
\end{eqnarray}
\end{subequations}
where, again, $\mathcal{M}=r_+^2/\sqrt{k}$. The first two terms of
metric (\ref{eq:ggwebs}) describe the extremal black string,
$\mathcal{Q}^2=\mathcal{M}^2$, in null coordinates
$v=(\hat{t}+\hat{x})/2,u=\hat{t}-\hat{x}$. In fact, using an
appropriate parameterization for $\bm{k}^{\mathrm{s}}$ the above
metric allows the following representation
\begin{equation}
g_{\mu\nu}=g^{\mathrm{e}}_{\mu\nu}+\frac{\mathcal{M}}{\hat{r}}
\left(1-\frac{\mathcal{M}}{\hat{r}}\right)^{-2}
k^{\mathrm{s}}_{\mu}k^{\mathrm{s}}_\nu, \label{eq:K-S}
\end{equation}
where $\bm{g}^{\mathrm{e}}$ is the metric of the extremal black
string. Since $\bm{k}^{\mathrm{s}}$ is a null Killing field it is
also geodesic, hence the above expression represents a generalized
Kerr-Schild transformation of the extremal black string. In other
words, the string background (\ref{eq:gwebs}) describes an exact
gravitational wave propagating along the extremal black string
\cite{Garfinkle:1992zj}.

In Ref.~\cite{Horowitz:1993} it was shown that the extremal BTZ
black hole is T-dual to this gravitational wave (see also
\cite{Ginsparg:1992af,Quevedo:1992ts}). Here we have made the
self-T-duality of this wave-like solution explicit, and showed that
this is realized along the direction
\begin{equation}
\bm{\partial_x}=\frac{1}{2}\bm{\partial_v}-\bm{\partial_u}.
\label{eq:xgwebs}
\end{equation}
The next case that requires to be studied would be that for which
all integration constants vanish.

\subsection{\label{subsec:fs}Case $r_+=0=r_-$: Flat space with linear
dilaton}

This case turns out to be the simplest one. By using the following
coordinates
\begin{equation}
(t,r,\varphi)\mapsto
\left(\bar{t}=at/\sqrt{k},\,\bar{r}=\sqrt{k}\ln(r/a),\,x=a\varphi\right),
\end{equation}
background (\ref{eq:sol}) simply becomes flat space with a linear
dilaton and vanishing axion; namely
\begin{subequations}
\label{eq:fs}
\begin{eqnarray}
\bm{g} &=&-\bm{d\bar{t}}^{2}+\bm{d\bar{r}}^{2}+\bm{dx}^{2}, \\
\Phi &=&-\frac{\bar{r}}{\sqrt{k}}.
\end{eqnarray}
\end{subequations}
In this case the self-T-duality along the direction
$\bm{\partial_x}$ is explicitly manifest. It is interesting to
remark that this background is the asymptotic geometry of the two
previously studied cases when their coordinate $\hat{r}$ goes to
infinity.

Finally, let us briefly discuss the cases where $|J|>M\protect\sqrt{k}$.

\subsection{\label{subsec:JbM}Case $|J|>M\protect\sqrt{k}$}

So far, we have examined those cases for which the horizon radius
$r_{\pm}$ turns out to be defined, and this actually occurs for
values such that $|J|\leq{M}\sqrt{k}$. The reason for this is that
for $|J|>M\sqrt{k}$ the geometry (\ref{eq:gcbs}) presents a naked
singularity at $r=0$. This is similar to the case of the BTZ
geometry. However, even though we focus our attention to those
geometries where no such singularities exist, we find illustrative
to discuss one particular case of that sort here. As it was
previously pointed out, for these values of the integration
constants the case with $J=0$ is special since it allows the
existence of hypersurface-orthogonal Killing fields in contrast to
the case $J\neq0$ where the existence of these fields is forbidden.
Let us describe this case in some detail. For vanishing $J$ the
condition $|J|>M\sqrt{k}$ implies that $M$ is negative. Then, by
choosing the following coordinates
\begin{eqnarray}
&&(t,r,\varphi )\mapsto \nonumber \\
&&\textstyle\left(\bar{t}=at/\sqrt{k},
\,\rho=\sqrt{k}\,\mathrm{arcsinh}({r}/{\sqrt{-Mk}}),
\,x=a\varphi\right),\qquad~
\end{eqnarray}
the background (\ref{eq:sol}) takes the form
\begin{subequations}
\label{eq:J0M<0}
\begin{eqnarray}
\bm{g}
&=&-\coth^{2}\left(\frac{\rho}{\sqrt{k}}\right)\bm{d\bar{t}}^{2}
+\bm{d\rho}^{2}+\bm{dx}^{2}, \\
\Phi   &=&-\ln\left[\frac{\sqrt{-Mk}}{a}
\sinh\left(\frac{\rho}{\sqrt{k}}\right)\right],
\end{eqnarray}
\end{subequations}
with a vanishing Kalb--Ramond field. This geometry turns out to be
dual to the Witten 2D black hole times the real line
(\ref{eq:Witten}), if the duality transformation is thought of to be
performed along the timelike direction $\bar{t}$. Timelike T-duality
was discussed in Ref.~\cite{Welch:1994qm} within the context of
three-dimensional string theory, and it was shown to relate positive
mass solutions to singular analogs of negative mass. Here, we are
only considering standard spacelike duality transformations instead.
Actually, applying T-duality in the spacelike $x$-direction one
verifies that the background above remains invariant. It does
represent a self-dual background describing a naked singularity.

Let us also notice here that a double Wick rotation can be
performed, namely $\bar{t}\rightarrow{i}\theta$,
$x\rightarrow{i}\tau$, and then used to show that the metric takes
the form
\begin{equation}
\bm{g}^{\mathrm{t}}=-\bm{d\tau}^{2}+\bm{d\rho}^{2}
+\coth^{2}\left(\frac{\rho}{\sqrt{k}}\right)\bm{d\theta}^{2}.
\label{eq:trumpet}
\end{equation}
This geometry is also a solution of the low energy string equations,
and corresponds to the product between the time direction $\tau$ and
the often called \emph{trumpet geometry}, which turns out to be the
T-dual to the Witten cigar \cite{Dijkgraaf:1991ba}, i.e.\ this is
dual of the Euclidean version of the two-dimensional black hole. In
fact, by performing T-duality in the $\theta$-direction one gets
\begin{equation}
\widetilde{\bm{g}}^{\mathrm{t}}=-\bm{d\tau}^{2}+\bm{d\rho}^{2}
+\tanh^{2}\left(\frac{\rho}{\sqrt{k}}\right)\bm{d\theta}^{2},
\label{eq:cigar}
\end{equation}
which is the product between the time direction $\tau$ and the
Witten cigar. Since the function $\tanh^{2}({\rho}/{\sqrt{k}})$
vanishes at $\rho=0$, the $\theta$-direction finds a fixed point at
the origin, where its T-duality transformation is not actually
defined; this explains the divergence that its dual
(\ref{eq:trumpet}) develops at $\rho=0$. Notice that this is analog
to the quoted example of the three-dimensional Minkowski space
$\bm{\eta}$ expressed in polar coordinates ($\tau,\rho,\theta$),
which, once dualized along the $\theta$-direction, yields a dual
$\widetilde{\bm{\eta}}$ that develops a singularity at the origin
$\rho=0$ \cite{Horowitz:1994ei}. It is clear from
Eq.~(\ref{eq:cigar}) that performing the double Wick rotation
backwards one ends up with the Witten black hole times the line
(\ref{eq:Witten}). The trumpet geometry (\ref{eq:trumpet}) is also
T-dual to $AdS_{3}$ space, as it was studied in \cite{Horowitz:1993}
as a particular case of the BTZ geometry; and more important for us
is that (\ref{eq:J0M<0}) results self-dual as well. All this
\emph{cascade of dualities} turns out to be interesting and, indeed,
suggestive. First, we should point out that, from the viewpoint of
the CFT worldsheet formulation of string theory, both backgrounds
(\ref{eq:trumpet}) and (\ref{eq:cigar}) are completely equivalent.
The CFT involved turns out to be the $SL(2,\mathbb{R})_{k}/U(1)$
WZNW model, which presents a $SL(2,\mathbb{R})_{k} \times
SL(2,\mathbb{R})_{k}$ symmetry group, and the duality transformation
that translates one target-space picture into the other simply
corresponds to changing the sign in one of the currents $J^{a}$ that
generate one of the two $SL(2,\mathbb{R})$'s factors (let us say the
right-handed factor). This also resembles the symmetry under
dualizing the radius as $R\rightarrow{R}^{-1}$ in the
compactification of the free boson, which represents the
prototypical example to discuss the T-duality. Regarding the CFT
description of the T-duality, it would be certainly interesting to
fully understand how the duality symmetry connecting both
(\ref{eq:trumpet}) and (\ref{eq:cigar}), and the fact that T-duality
(though in a different direction) also connects the trumpet geometry
to the $AdS_{3}$ space. Actually, the CFT description of $AdS_3$
strings is closely related to that of the theory formulated on the
manifold (\ref{eq:cigar}) which, as mentioned, is the product of
time and the Euclidean 2D black hole. Algebraically, such relation
regards a natural realization of the symmetries that the worldsheet
theory presents; and having a geometrical picture of it provides a
way for working out the details of the connection existing between
both realizations.

\section{\label{sec:conclu}Conclusions}

In this paper, we studied the T-duality in three-dimensional bosonic
string theory from the viewpoint of the low energy effective action.
Such duality symmetry is known to manifest that different spacetime
configurations can be interpreted as two different regimes of the
same string background. Within the framework of the CFT description
of the theory, two models that are connected by T-duality are indeed
completely equivalent and consequently the two target-space
interpretations are equally valid. More generally, if the isometry
with respect to which one performs a given T-duality transformation
corresponds to a spacelike compact direction, then the original
solution and its dual correspond to the same conformal field theory
\cite{Rocek:1991ps}. More concisely, the CFT description of both
sides of the duality map are equivalent and, in terms of the stringy
description, this is typically realized by the interchange between
winding and Kaluza--Klein momenta in the compact direction.
According to this picture, those configurations that result to be
self-dual are such that this interchange of quantum numbers can be
realized on the same spacetime. These configurations are thought of
as gathering important properties of the string theory, in
particular in what respects to its symmetries.

Here, we investigated the simplest case of nontrivial self-T-dual
configurations and identified minimum criteria that yield a
classification of previously known exact solutions of
three-dimensional string theory. This amounts of imposing stringent
self-duality conditions, so that the duality transformation was
thought of to be performed along an isometry direction generated by
a Killing vector with constant norm. It is highly remarkable that
the pdes system resulting from the low energy string
equations become fully integrable in this case. The first
consequence of this fact is that the resulting solutions are
necessarily stationary, i.e.\ a Birkhoff' theorem of the sort of the
one proven in \cite{Ayon-Beato:2004if} for pure gravity is obeyed by
these configurations. An unusual fact we found through the
computations is that, in order to make integrability manifest, it
was necessary to take one of the gauge elections as imposed on the
string coupling: the dilaton. In spite of the fact that the above is
a nonstandard procedure it results fully justifiable at the local
level (as we show in Appendix \ref{app:gauge}).

Our main results can be stated as follows: for the case of
nontrivial string coupling, the lower energy string backgrounds with
a constant norm Killing field that are additionally self-T-dual are
uniquely described by a two-parametric class, including only three
nonsingular cases: the charged black string, the exact gravitational
wave propagating along the extremal black string, and flat space
with a linear dilaton. Besides, for a constant string coupling, the
only self-T-dual lower energy string background under the same
assumptions corresponds to the Coussaert--Henneaux spacetime. We
also discussed other cases and, along this work, we went through the
bestiary of three-dimensional string backgrounds. Actually, we
presented a survey of fixed points of T-duality transformations in
three-dimensional low energy effective bosonic string theory. We did
this by means of standard techniques for solving the equations of
motion of three-dimensional gravity models,
and the fact of having worked out a classification for the described
self-dual solutions in such a simple way provides an example about
how the techniques developed in Ref.~\cite{Ayon-Beato:2004if} were
suitable to be used within a more general context.

\begin{acknowledgments}
We thank A.~Anabal\'on, D.~Correa, J.~Edelstein, M.~Hassa\"{\i}ne,
C.~Mart\'{\i}nez, J.~Oliva, R.~Portugu\'{e}s, R.~Troncoso, and
J.~Zanelli for useful discussions. EA-B specially thanks to
H.~Garc\'{\i}a-Compe\'{a}n and C.~Soto-Campos for introducing him to
this subject. GG specially thanks the Centro de Estudios
Cient\'{\i}ficos (CECS), Valdivia, for the hospitality. This work
is partially supported by grants 1040921, 7040190, 1051064, and
1060831 from FONDECYT, grants CO1-41639 and CO2-44598 from CONACyT,
and by CONICET and Universidad de Buenos Aires. Institutional
support to the CECS from Empresas CMPC is gratefully acknowledged.
CECS is a Millennium Science Institute and is funded in part by
grants from Fundaci\'{o}n Andes and the Tinker Foundation. GG is
Member of the Consejo Nacional de Investigaciones Cient\'{\i}ficas y
T\'ecnicas (CONICET), Argentina, and Junior Associate to the Abdus
Salam International Centre for Theoretical Physics (ICTP), Italy.
\end{acknowledgments}

\appendix

\section{\label{app:SEqs}Low energy string equations for self-T-dual
backgrounds}

A low energy string background is determined by Eqs.~(\ref{eq:les}).
In three dimensions the Kalb--Ramond strength must be proportional
to the volume three-form
$\eta_{\alpha\beta\gamma}=\sqrt{-g}\epsilon_{\alpha\beta\gamma}$
($\epsilon_{tr\varphi}=+1$) and such proportionality is
straightforwardly fixed by Eq.~(\ref{eq:KR}) as being
\begin{equation}
H_{\alpha\beta\gamma}=c\,e^{2\Phi}\eta_{\alpha\beta\gamma},
\label{eq:solH}
\end{equation}
where $c$ is an integration constant related to the axion charge per
unit length. Using the above expression, the independent Einstein
equations for a geometric background of the form (\ref{eq:gexpl})
take the following form
\begin{subequations}
\label{eq:lesexpl}
\begin{eqnarray}
-\frac{NF}2(E_t{}^t-E_r{}^r-WE_\varphi{}^t) =
\Biggl(\frac{\dot{\Phi}}{N}\dot{\Biggr)~~}\!\!\!  \nonumber\\
{}+N^2F^2\left(\frac{\Phi^{\prime}}{N}\right)^{\prime}&=&0,
\label{eq:P..''} \\
& &  \nonumber \\
E_r{}^r+E_\varphi{}^\varphi+WE_\varphi{}^t =
\frac{e^{2\Phi}}{2N}\Biggl(
\frac{e^{-2\Phi} (F^{-1}\dot{)~}\!\!}N\dot{\Biggr)~~}\!\!\!\nonumber\\
{}-\frac{e^{-2\Phi}}{2N}\left(\frac{e^{6\Phi}
\left(e^{-4\Phi}N^2F\right)^{\prime}}N\right)^{\prime}  \nonumber\\
{}-4F(\Phi^{\prime})^2+c^2e^{4\Phi} &=&0,
\label{eq:F..''} \\
& &  \nonumber \\
-E_r{}^t = \frac{(F\Phi^{\prime}\dot{)~}\!\!}{N^2F^2}
+\left(\frac{\dot{\Phi}}{N^2F}\right)^{\prime}&=&0,
\label{eq:P.'} \\
& &  \nonumber \\
\frac{2Ne^{-2\Phi}}{a^2}E_\varphi{}^t =
\left(\frac{e^{-2\Phi}W^{\prime}}{N}\right)^{\prime}&=&0,
\label{eq:W''} \\
& &  \nonumber \\
-\frac{2Ne^{-2\Phi}}{a^2}E_\varphi{}^r =
\biggl(\frac{e^{-2\Phi}W^{\prime}}{N}\dot{\biggr)~~}\!\!\! &=&0,
\label{eq:W.'} \\
& &  \nonumber \\
-\frac{2e^{-4\Phi}}{a^2}(E_\varphi{}^\varphi+WE_\varphi{}^t) =
\left(\frac{e^{-2\Phi}W^{\prime}}N\right)^2-\frac{c^2}{a^2}
&=&0,\qquad~\label{eq:W'}
\end{eqnarray}
where $\dot{(\ldots)}$ and $(\ldots)^{\prime}$ denote derivatives
with respect to the coordinates $t$ and $r$, respectively, and
$E_{\alpha\beta}=0$ are the components of the Einstein equations
(\ref{eq:Ein}). The remaining dilaton equation (\ref{eq:dil}) is
given by
\begin{equation}
-\frac{e^{2\Phi}}N
\Biggl(\frac{e^{-2\Phi}\dot{\Phi}}{NF}\dot{\Biggr)~~} \!\!\!
+\frac{e^{2\Phi}}N\left(NFe^{-2\Phi}\Phi^{\prime}\right)^{\prime}-
\frac{c^2}2e^{4\Phi} +\frac2k=0.\label{eq:dila}
\end{equation}
\end{subequations}
From Eqs.~(\ref{eq:W''}) and (\ref{eq:W.'}) it is clear that the
quantity
\begin{equation}
J=\frac{a^3e^{-2\Phi}W^{\prime}}N, \label{eq:Wr}
\end{equation}
is an integration constant. The remaining equations determine the
form of $F(t,r)$, $N(t,r)$, and $W(t,r)$, while $\Phi(t,r)$ is fixed
by choosing an appropriate coordinate, as it is justified in the
Appendix B.

\section{\label{app:gauge}Justifying gauge elections}

Something that can be found puzzling is the fact that the last gauge
election on the cases $\nu^2\neq0$ of Sec.~\ref{sec:Birk} were taken
on the dilaton and not on the metric functions as usual. However,
this choice is fully consistent with the previous ones which allow
to write the metric as in Eq.~(\ref{eq:gexpl}). In order to avoid
any confusion we dedicate this appendix to justify this item. The
key point here is that after fixing $g_{tr}=0=g_{\varphi{r}}$,
metric (\ref{eq:gexpl}) presents a residual symmetry, it is
form-invariant under the coordinate transformation
\begin{equation}
(t,r,\varphi)\mapsto
\left(\check{t}=f_1(t,r),\,\check{r}=f_2(t,r),\,
\check{\varphi}=\varphi+f_3(t,r)\right), \label{eq:res}
\end{equation}
together with the redefinitions
\begin{eqnarray}
\check{F} &=&
F\left((f_2^\prime)^2-\frac{(\dot{f_2})^2}{N^2F^2}\right),
\label{eq:Fn} \\
\check{N} &=& N\frac{f_2^\prime}{\dot{f_1}}
\left((f_2^\prime)^2-\frac{(\dot{f_2})^2}{N^2F^2}\right)^{-1}, \\
\check{W} &=& \frac{W-\dot{f_3}}{\dot{f_1}},
\end{eqnarray}
where the functions $f_i$, $i=1,2,3$, obey the two
conditions
\begin{eqnarray}
f_1^{\prime}f_2^{\prime}-\frac{\dot{f_1}\dot{f_2}}{N^2F^2} &=& 0,
\label{eq:c1}\\
\dot{f_1}f_3^{\prime}-f_1^{\prime}(\dot{f_3}-W) &=& 0,\label{eq:c2}
\end{eqnarray}
related to the fact that the transformation (\ref{eq:res}) respects
the gauge election, i.e.
$g_{\check{t}\check{r}}=0=g_{\check{\varphi}\check{r}}$. So, the
remaining gauge choice just corresponds to fix one of the above
functions. Hence, in order to recover, for example, the gauge
election (\ref{eq:gaugePhi(r)}) we just need to make a
transformation of the type discussed above, with $f_2=ae^{-\Phi}$,
and where the functions $f_1$ and $f_3$ are obtained from the linear
first order pdes
\begin{eqnarray}
\left(\partial_t-\frac{\Phi^{\prime}N^2F^2}{\dot{\Phi}}\partial_r\right)
f_1&=&0,\\
\left(\partial_t-\frac{\Phi^{\prime}N^2F^2}{\dot{\Phi}}\partial_r\right)
f_3&=&W.
\end{eqnarray}
On the other hand, the gauge election (\ref{eq:gaugePhi(t)}) is
obtained by means of a similar transformation, where this time
$f_1=ae^{-\Phi}$ and where the functions $f_2$ and $f_3$ come from
solving the linear first order pdes
\begin{eqnarray}
\left(\partial_t-\frac{\Phi^{\prime}N^2F^2}{\dot{\Phi}}\partial_r\right)
f_2&=&0,\\
\left(\partial_t-\frac{\dot{\Phi}}{\Phi^{\prime}}\partial_r\right)
f_3&=&W.
\end{eqnarray}
Finally, for the gauge choice of Subsec.~\ref{subsec:null} we
substitute $\check{F}=1$ in Eq.~(\ref{eq:Fn}) which provides a third
condition on the functions $f_i$, $i=1,2,3$, additional to the
conditions (\ref{eq:c1}) and (\ref{eq:c2}), and additionally
determines the corresponding transformation. The solutions of all
the previous first order pdes can be found integrating their
corresponding characteristic ordinary systems. This guarantees the
existence of the related coordinate systems.

\section{\label{app:par}On self-T-dual directions with nonconstant norm}

Now, let us comment on the existence of more general examples than,
while still satisfying the requirements for self-T-duality, do not
necessarily present a constant norm Killing vector along the
isometry direction where the T-duality is being performed. We
explore this possibility within a simple set-up according to which
the Kalb--Ramond field $B_{\mu\nu}$ is set to zero and the metric
and dilaton acquire the following form
\begin{subequations}\label{eq:s-dnc}
\begin{eqnarray}
\bm{g}&=&e^{2\xi(t,y)}(-\bm{dt}^{2}+\bm{dy}^{2})
        +e^{2\Psi(t,y)}\bm{dx}^{2},\\
\Phi  &=&\Phi(t,y),
\end{eqnarray}
\end{subequations}
for a pair of differentiable functions $\xi$ and $\Psi$ which, like
the dilaton $\Phi$, are assumed to depend only on the coordinates
$t$ and $y$.

Thus, the idea is to find a configuration that could still be
self-T-dual along the direction $\bm{\partial_x}$ even though the
function $\frac{1}{2}\ln g_{xx}=\Psi(t,y)$ is nonconstant. This
would be possible due to the fact that a self-T-dual solution is one
that, after performing the T-duality transformation, recovers its
original form up to diffeomorphisms and gauge transformations.
Hence, a given configuration (\ref{eq:s-dnc}) is self-dual if there
exists a diffeomorphism
\begin{equation}
\ast\!: (t,y) \mapsto \left(T(t,y),Y(t,y)\right),
\end{equation}
that leaves the two-dimensional metric block
\[
\bm{g}^{(2)}=e^{2\xi(t,y)}(-\bm{dt}^{2}+\bm{dy}^{2}),
\]
invariant, and for which the conditions
\begin{subequations}\label{eq:sdc}
\begin{eqnarray}
\widetilde{\Psi}(t,y)&=&-\Psi(t,y)=\Psi(T(t,y),Y(t,y)),
                        \label{eq:sdc1}\\
\widetilde{\Phi}(t,y)&=&\Phi(t,y)-\Psi(t,y)=\Phi(T(t,y),Y(t,y)),\qquad~
                        \label{eq:sdc2}
\end{eqnarray}
\end{subequations}
are obeyed. These functional conditions does not
look a priory so restrictive; in particular, if the coordinate
transformation satisfy to be an involution, $\ast^{2}=1$, then the
first condition (\ref{eq:sdc1}) is a consequence of the second one
(\ref{eq:sdc2}), meaning that it is simply requesting that the
corresponding metric function is the non-invariant part of the
dilaton under the action of $*$, so that $\Psi(t,y)$ is odd with
respect to it.

It is instructive to think in the concise example where $\xi(t,y)$
turns out to be symmetric under parity transformation $\ast\!\!: y
\mapsto -y$ while the function $\Psi(t,y)$ is assumed to agrees the
odd part of the dilaton, namely $\Psi(t,y)=\Phi(t,y)-\Phi(t,-y)$.
Notice that such an ansatz would lead to a self-T-dual configuration
even for a nonconstant $\Psi(t,y)$ satisfying these requirements. In
fact, by applying T-duality along the direction $x$, one would
obtain
\begin{subequations}
\begin{eqnarray*}
\widetilde{\bm{g}}&=&e^{2\xi(t,y)}(-\bm{dt}^{2}+\bm{dy}^{2})
                    +e^{-2\Psi(t,y)}\bm{dx}^{2},\\
\widetilde{\Phi}  &=&\Phi(t,y)-\Psi(t,y),
\end{eqnarray*}
\end{subequations}
which means that $\Psi(t,y)$ is odd under reversing the sign of $y$.
Now, the requirements mentioned above would yield the following dual
configurations
\begin{eqnarray*}
\widetilde{\bm{g}}&=&e^{2\xi(t,-y)}(-\bm{dt}^{2}+(-1)^{2}\bm{dy}^{2})
                    +e^{2\Psi(t,-y)}\bm{dx}^{2}, \\
\widetilde{\Phi}  &=&\Phi(t,-y).
\end{eqnarray*}
Hence, by simply renaming coordinates as $y \mapsto Y=-y$, we would
certainly reobtain the original configuration. This can be
straightforwardly extended to the generic form $\ast\!\!\!: (t,y)
\mapsto (T(t,y),Y(t,y))$ we were discussing above.

This kind of construction would provide concrete examples of
self-dual backgrounds only if the conditions (\ref{eq:sdc}) are
compatible with the low energy string equations. Nevertheless, the
straightforward efforts to find explicit non-trivial examples do not
prove to succeed. In fact, the stationary branch related to the
configuration (\ref{eq:s-dnc}), when there is no dependence on time,
turns out to be fully determined from the low energy string
equations by using a different gauge election. The resulting
stationary configuration is only self-dual if $\Psi(y)=0$, i.e.\ for
a constant norm self-dual direction.

\end{document}